\documentclass{emulateapj}

\usepackage{amssymb}
\usepackage{amsmath}
\usepackage{natbib}
\usepackage{txfonts}
\usepackage[colorlinks=true,citecolor=blue,linkcolor=blue]{hyperref}
\usepackage{microtype}
\usepackage{color}
\usepackage{graphicx}
\usepackage{epstopdf}

\def\apj{ApJ}
\def\apjl{ApJL}
\def\apjs{ApJS}

\def\aap{A\&A}

\def\nice{1}
\def\roma{2}
\def\icranet{3}
\def\icranetrio{4}

\shorttitle{Radial stability in stratified stars}
\shortauthors{Pereira and Rueda}

\begin{document}

\title{Radial stability in stratified stars}

\author{Jonas P.~Pereira\altaffilmark{\nice,\roma} and Jorge A. Rueda\altaffilmark{\roma,\icranet,\icranetrio}}

\altaffiltext{\nice}{Universit\'e de Nice Sophia Antipolis,
										CEDEX 2, Grand Ch\^ateau Parc Valrose,
										BP 2135, 06103 Nice,
										France}
\altaffiltext{\roma}{Dipartimento di Fisica and ICRA,
                     Sapienza Universit\`a di Roma,
                     P.le Aldo Moro 5,
                     I--00185 Rome,
                     Italy}

\altaffiltext{\icranet}{ICRANet,
                     P.zza della Repubblica 10,
                     I--65122 Pescara,
                     Italy}

\altaffiltext{\icranetrio}{ICRANet-Rio,
											Centro Brasileiro de Pesquisas F\'isicas,
											Rua Dr.~Xavier Sigaud 150,
											Rio de Janeiro, RJ, 22290--180,
											Brazil}

\begin{abstract}
We formulate within a generalized distributional approach the treatment of the stability against radial perturbations for both neutral and charged stratified stars in Newtonian and Einstein's gravity. We obtain from this approach the boundary conditions connecting two any phases within a star and underline its relevance for realistic models of compact stars with phase transitions, owing to the modification of the star's set of eigenmodes with respect to the continuous case.
\end{abstract}


\altaffiltext{}{jonaspedro.pereira@icranet.org; jorge.rueda@icra.it}

\maketitle

\section{Introduction}

There is theoretical evidence that compact stars, such as neutron stars, are made up of several matter phases \citep{1986bhwd.book.....S}. This is a consequence of the astonishingly high density excursion they could attain in their inner regions. For instance, a neutron star is in general thought as composed by at least to different regions: the crust and the core. Starting from very low densities of a few grams per cubic centimeter close to their surfaces, and up to densities of the order of the nuclear saturation value, $\rho_{\rm nuc}\approx 2.7\times 10^{14}$~g~cm$^{-3}$, the crust of a neutron star is thought to be in a \emph{solid-like} state. The core, with densities that might overcome by orders of magnitude $\rho_{\rm nuc}$, is instead thought to be in a \emph{liquid-like} state. The details of the treatment of the thermodynamic transition (e.g.~Maxwell or Gibbs phase construction), as well as the conditions on density and pressure at which such a transition occurs, are still a matter of debate. The application of the Gibbs construction with more than one conserved charge (e.g.~baryon and electric) leads to the appearance of mixed phases, in between of the pure phases, with an equilibrium pressure that varies with the density, leading to a spatially extended phase-transition region of non-negligible thickness with respect to the star's radius \citep{1992PhRvD..46.1274G,1995PhRvC..52.2250G,1997PhRvC..56.2858C,1999PhRvC..60b5803G,2000PhRvC..62b5804C,2001PhR...342..393G}. On the contrary, in the traditional Maxwell construction the phases are in ``contact'' each other. It is worth mentioning that in these treatments the pure phases are subjected to the condition of local charge neutrality, and so they do not account for the possible interior Coulomb fields. Indeed, the complete equilibrium of the multicomponent fluid in the cores of compact stars needs the presence of a Coulomb potential formed by electric charge separation due to gravito-polarization effects \citep{Rotondo:2011rj,Rueda:2011nq}, favoring a sharp core-crust transition that ensures the global, but not the local, charge neutrality \citep[see][and references therein]{2012NuPhA.883....1B,2014NuPhA.921...33B}. Besides the core-crust transition, additional phase-transitions as the ones allowed by Quantum Chromo-Dynamics (QCD) could occur within the core of the star itself \citep[see, e.g.,][and references therein]{1996csnp.book.....G}. From all the above we can conclude that ultradense stars such as neutron stars shows necessarily a nontrivial stratification. Between any two phases, which can be very different, it is reasonable to investigate the situation where some quantities are discontinuous, e.g.~the energy-density and the pressure. Such discontinuities can be harnessed by appropriate surface tensions. These surface quantities influence the stability of a system adding new boundary conditions to the problem which, as we shall show here, modify the set of eigenfrequencies and eigenmodes of a star.

In this work we analyze the problem of perturbations in systems constituted of various phases that are split by surfaces that host nontrivial degrees of freedom. This analysis is thought to be a generalization of the treatment for continuous systems \citep[see][and references therein, for a comprehensive analysis of properties, types and stability of continuous anisotropic fluids also in presence of radiation and heat flux]{1997PhR...286...53H}. By investigating the dynamics of perturbations, we are automatically probing the stability of systems. We shall trammel ourselves to the simplest possible case: spherically symmetric extended bodies where radial perturbations take place. In order to model the problem, we shall assume these surfaces of discontinuity separating two arbitrary phases are very thin and a generalized distributional approach \citep{2004reto.book.....P, 1982JPhA...15..381R} shall be adopted. We start our analysis in the Newtonian case in order to gain intuition of the relevant aspects of the problem, and finally we generalize it to general relativity. Our purpose is solely to expound the problem and seek to solve it as generically as we can. Our analysis is far from being complete and has to be considered as a first step towards deeper investigations and scrutinies of specific cases shall be the object of posterior studies.

We show in this work that phase transitions in the presence of surface degrees of freedom can be enclosed into additional boundary conditions to the problem. Besides, our formalism tells us that such boundary conditions are only self-consistent when the set of eigenfrequencies of the perturbation modes is related to the global system, not with individual phases. This is consistent with the well-known results from coupled springs, where there are only global frequencies. The presence of further boundary conditions naturally modify the possible set of eigenfrequencies, since we are inserting further restrictive aspects to the physical oscillation modes. Therefore, measurements on the pulsation modes in a star could tell us very precisely about its internal structure, being a sort of fingerprint, which could help to understand better the nature of these systems.

The article is organized as follows. In section~\ref{sec:2} we formulate the problem of radial perturbations of stars with interfaces in the classical Newtonian case. For the sake of exemplification, we compute the specific solution of the stability equation in section~\ref{sec:3} for the simple case of incompressible stars, scrutinizing which circumstances deliver us enough arbitrary constants to fix additional boundary conditions. In section~\ref{sec:5} we give the formulation in the case of systems with electric charge but still within Newtonian gravity. The formulation in the case of general relativity in the neutral case is presented in section~\ref{sec:6}, while in section~\ref{sec:7} we extend the general relativistic treatment to systems endowed with electric charge. Finally, in section~\ref{sec:8} we summarize and discuss the main results of this work. We use units such that $c=G=1$ and the metric signature $-2$ throughout the article, unless otherwise stated.

\section{Stability of classical systems with phase transitions}\label{sec:2}

Assume a continuous classical astrophysical system with spherical symmetry. When its volume elements are perturbed radially, it is well-known \citep[see, e.g.,][]{1986bhwd.book.....S} that the evolution of perturbations of the form
\begin{equation}\label{eq:xi}
\tilde\xi(r,t)=\xi(r)e^{i \omega t},
\end{equation}
with $\omega$ an arbitrary constant, is described by
\begin{equation}
\frac{d}{dr}\left[\Gamma P\frac{1}{r^2}\frac{d(r^2\xi)}{dr} \right]-\frac{4}{r}\frac{dP}{dr}\xi +\omega^2\rho\xi=0\label{perteq},
\end{equation}
where $P(r)$ is the pressure of the background system under hydrostatic equilibrium and
\begin{equation}
\Gamma\doteq \frac{\rho}{P}\frac{\partial P}{\partial \rho} \label{Gamma},
\end{equation}
with $\rho(r)$ the mass density of the system. Formulated in this way, we have at hands an eigenvalue problem. For the case of continuous systems, the boundary conditions to be supplemented to Eq.~(\ref{perteq}) are quite simple. They are directly related to the spherical symmetry of the system, as well as to the vanishing of its pressure on its border even in the presence of perturbations (other situations where a surface tension be present could also be envisaged and we shall attempt to elaborate on them in the sequel). In other words, we impose that
\begin{equation}
\xi(0)=0,\;\;\mbox{and}\;\;\xi(R_s)=\mbox{finite}\label{boundarycont},
\end{equation}
where $R_s$ was defined as the radius of the star, such that $P(R_s)=0$. For further details about these boundary conditions \citep{1986bhwd.book.....S}. Eq.~(\ref{perteq}) supplemented with Eq.~(\ref{boundarycont}) constitutes a Sturm-Liouville problem, where the aspects of its solutions are already known. Concerning the eigenfrequencies, $\omega^2$, they are all real and form a discrete hierarchical set. When one seeks for stable solutions to Eq.~(\ref{perteq}), one seeks for solutions with positive $\omega^2$, specially for its fundamental mode. As it can be seen from Eq.~(\ref{eq:xi}), negative values of $\omega^2$ indicate instabilities in the assumed background system, which leads to the conclusion that they do not linger on in time. They would either implode or explode.

We now turn to the more involved problem of permitting the system to be stratified and harboring surface degrees of freedom on the interface of two given phases.
Such degrees of freedom have themselves a dynamics, described generically by the thin-shell formalism or Darmois-Israel formalism \citep{1966NCimB..44....1I,1967NCimB..48..463I,2005CQGra..22.4869L,2004reto.book.....P}. We are here however particularly interested in another aspect of the problem, namely to understand the role such degrees of freedom play on the stability of the system when its parts (defined naturally by the hypersurface that hosts the aforesaid degrees of freedom) are perturbed. Therefore, before anything, it is assumed that for being meaningful to talk about this scenario, one has that the hypersurfaces of discontinuity themselves are stable \citep[see][for further details]{2014arXiv1412.1848P}. If this is not the case, any displacement of the hypersurface of discontinuity would trigger a cataclysmic set of events that would have as a result the disruption of the system.

One expects that the stratified problem could be accounted for additional boundary conditions to the system. The reason for this is that the perturbations in the upper and lower regions with respect to a given surface of discontinuity would be described by the same physics [e.g.~Eq.~(\ref{perteq})], as well as the totality of the matches. The only missing points would be their connection (allowing combinations of solutions to be also solutions to the physical equations involved) and generalization by means of surface quantities. For example, the existence of a surface tension would account to an extra surface force term. The same ensues with the presence of a surface mass (enclosed by a surface mass density) and the associated presence of a surface gravitational force. Therefore, in order to properly describe stratified systems, one must make use of distributions \citep{2004reto.book.....P, 1982JPhA...15..381R}.

We proceed now with the distributional generalization of the equations describing continuous fluids under gravitational fields. Assume that a surface harboring surface degrees of freedom of a system in equilibrium is at $r=R$. The first equation to be generalized in terms of distributions in this case is the equation of hydrostatic equilibrium. This should now read
\begin{equation}
\frac{d P}{dr} + \rho g(r) - \frac{2{\cal P}}{R}\delta(r-R)=0 \label{hydrostaticeq},
\end{equation}
where $g(r)$ has been defined as the norm of the gravitational field, solution to (the distributional)
Poisson's equation $\nabla \cdot \vec{g}= -4\pi G \rho$, $\vec{g}=-g(r)\hat{r}$, that can always be written as
\begin{equation}
g(r)=\frac{GM(r)}{r^2},\;\;M(r)\doteq 4\pi\int^{r}_0\rho(\bar{r})\bar{r}^2d\bar{r} \label{gravfield}.
\end{equation}
Besides, ${\cal P}$ in Eq.~(\ref{hydrostaticeq}) stands for the surface tension \citep{2013ConPh..54...60P} on the surface of discontinuity at $r=R$.
The expression on the aforesaid equation is {the result of} restoring surface forces (this is the reason they have an opposite direction
to the pressure gradient) at a small surface area (the factor $2$ comes from the principal curvatures in a surface element, which in the spherically symmetric case are equal). The gravitational force from the sheet of mass at $r=R$ is naturally incorporated on the distributional definitions of $\rho$ and $g(r)$ as we shall show in the sequel. The mass density, on the other hand, must be expressed as
\begin{equation}
\rho(r)=\rho^-(r)\theta(R-r) + \rho^+(r)\theta(r-R) + \sigma \delta(r-R)\label{rhodistr},
\end{equation}
with $\theta(r-R)$ the Heaviside function, whose derivative is the Dirac delta function $\delta(r-R)$ {in the sense of distributions}. From the existence of a surface mass density, it can be checked that the gravitational field is discontinuous at a given surface of discontinuity of the system (at $R$), i.e.
\begin{equation}
[g(R)]^+_-=4\pi G \sigma,\label{gravfieldjump}
\end{equation}
where we have introduced the convention $[A]^{+}_{-}\doteq A^{+}-A^{-}$, the jump of $A(r)$ across $r=R$. Therefore, $g(r)$ could be represented distributionally as
\begin{equation}
g=g^+(r)\theta(r-R)+g^-(r)\theta(R-r) \label{gravfielddistr}.
\end{equation}
The above equation means that the associated distributional gravitational potential $\phi(r)$ ($\vec{g}= -\nabla \phi$, $\vec{g}= - g\hat{r}$) is always a continuous function, though not differentiable at a surface of discontinuity.

Besides, we will assume that also the pressure can be discontinuous at $r=R$, being hence written as
\begin{equation}
P(r)=P^-(r)\theta(R-r) + P^+(r)\theta(r-R)\label{pdistr}.
\end{equation}
The reason of why this is so will be clarified when we deal with our original problem in the scope of general relativity. The heuristic argument corroborating the validity of Eq.~(\ref{pdistr}) is that it is meaningless to colligate a surface term to the radial pressure, since its associated force would necessarily be normal to it and therefore would not lie on the surface. Only tangential pressures should have tied within their surface terms. From Eqs.~(\ref{pdistr}), (\ref{rhodistr}), (\ref{gravfield}) (\ref{gravfieldjump}) and (\ref{gravfielddistr}), we have
\begin{equation}
{\cal P} = \frac{R}{2}[P(R)]^+_- + \frac{G}{16\pi R^3}[M^2(R)]^+_- \label{surfacetension}
\end{equation}
and
\begin{equation}
\frac{dP^{\pm}}{dr} + \rho^{\pm}(r)g^{\pm}(r)=0\label{hydrostaticeqdistr}.
\end{equation}

The arithmetic average present in Eq.~(\ref{surfacetension}) [see the definition of $g(r)$ and the value of $\sigma$] is a general consequence of the product of delta functions with Heaviside ones in the generalized sense of distributions \citep{1982JPhA...15..381R,1982JPhA...15.3915R}. Notice from Eq.~(\ref{surfacetension}) that its first term is the known Young-Laplace equation for spherical surfaces at equilibrium \citep[see, e.g.,]{2003EJPh...24..159R,2013ConPh..54...60P}, where just geometric aspects are taken into account for the surface tension. Its second term, though, is the gravitational surface tension, uniquely due to the non-zero surface mass. If the surface tension were null, the pressure jump could not be arbitrary, but proportional to the surface mass density and must be a monotonically decreasing function of the radial coordinate [see Eqs.~(\ref{gravfieldjump}) and (\ref{surfacetension})]. From Eq.~(\ref{surfacetension}), one sees further that the force per unit area associated with the surface tension is exactly the one necessary to counterbalance both the forces coming from the pressure gradient at $R$ and the surface gravitational force, as it should be.

One sees that the above procedure generalizes our notion of hydrostatic equilibrium in each phase the stratified system has [see Eq.~(\ref{hydrostaticeqdistr})] and automatically gives the surface tension at $R$ that guarantees the hydrostatic equilibrium for arbitrary pressure jumps and surface masses. We will
keep the same philosophy now concerning the generalization of Eq.~(\ref{perteq}). From our generalized hydrostatic equilibrium equation, we have that
an important term for the deduction of the equation governing radial perturbations would be the application of the Lagrangian operator $\Delta$ [$\Delta A \doteq A(t, r+\tilde\xi) - A_0(t,r)$, $A_0$ and $A$ being a physical quantity in the equilibrium and perturbed cases, respectively] on the surface force in
Eq.~(\ref{hydrostaticeq}), i.e.~[see Eq.~(\ref{masterpertdistrib})]
\begin{equation}
\Delta \left[ \frac{{\cal P}}{R} \delta(r-R)\right]\doteq  \frac{\Delta {\cal P}}{R}\delta(r-R) - \frac{{\cal P}\tilde\xi}{R^2} \delta(r-R)\label{lagrsurftension},
\end{equation}
since $\Delta \delta(r-R) = 0$ and $\Delta R= \tilde\xi$. Now we assume that ${\cal P}={\cal P} (\sigma)$. This means that we are endowing the fluid at the surface of discontinuity with adiabatic properties and the underlying microphysics is not contemplated in this procedure. For continuous media, the total mass in the interface of two phases is generally not a constant. This means that mass fluxes are allowed to take place. This generically would render the mass of each phase not constant, an aspect not taken into account in Eq.~(\ref{perteq}). Nevertheless, if the displacements of the surface of discontinuity are small and oscillatory, we have that on average the masses on each phase are conserved (here it becomes clear why the surface of discontinuity should be stable).
For adiabatic processes, we have
\begin{equation}
\Delta {\cal P}=\eta^2\Delta \sigma,\;\; \eta^2\doteq \frac{\partial {\cal P}}{\partial \sigma}\label{eta2},
\end{equation}
with $\eta^2$ the square of the speed of the sound in the fluid at the surface of discontinuity. The missing term $\Delta \sigma$ can be found via the thin-shell formalism when the classical limit is taken there. Generically, in the static and spherically symmetric case, $\sigma$ can be written as \citep{2005CQGra..22.4869L}
\begin{equation}
\sigma = -\frac{c^2}{4\pi G R}[e^{-\beta(R)}]^+_-\label{sigmagr},
\end{equation}
with the classical limit $\beta(r)\approx GM(r)/(rc^2)\ll 1$. It is easy to check that Eq.~(\ref{sigmagr}) reduces to Eq.~(\ref{gravfieldjump}) in the aforementioned limit. When perturbed, it can be shown that $\beta \rightarrow \beta + \delta \beta$, with $\delta \beta = -4\pi G r \rho_0 e^{2\beta_0}\tilde\xi/c^2$ \citep{1973grav.book.....M}, $\rho_0$ here meaning the mass density in the hydrostatic background solution. Hence,
\begin{equation}
\Delta \sigma = \delta \sigma + \sigma'_0 \tilde\xi = -[\rho_0\tilde\xi]^+_- + \sigma'_0\tilde\xi \label{Deltasigma},
\end{equation}
where we also considered $\sigma_0$ the background solution [Eq.~(\ref{gravfieldjump})].

Another simpler way of obtaining Eq.~(\ref{Deltasigma}) would be through the dynamics of $\delta g$ [$\delta \vec{g}$= -($\delta g) \hat r$].
In the spherically symmetric case we have $\delta g = -4\pi G \rho \tilde\xi$ \citep{1986bhwd.book.....S}, and since $\delta g \doteq g-g_0$,
with $g_0$ the norm of the gravitational field without the perturbation $\tilde \xi$, from Eq.~(\ref{gravfieldjump}), we finally obtain
\begin{equation}
\Delta \sigma = \frac{[\delta g(R)]^+_-}{4\pi G} + \sigma'_0(R)\tilde\xi = \sigma'_0(R)\tilde\xi - [\rho_0\tilde\xi]^+_-\label{Deltasigmasimpler}.
\end{equation}
Besides, from Eqs.~(\ref{gravfield}) and (\ref{gravfieldjump}), for the case when the jump of $\tilde\xi$ is null at a surface of discontinuity (that shall be justified in the sequel), one shows that the above equation can be further simplified to
\begin{equation}
\Delta \sigma= -\frac{2}{r}\sigma \tilde{\xi}\label{Deltasigmasimpl}.
\end{equation}
Now we show the general equation governing the propagation of radial perturbations. For $P$ defined as in Eq.~(\ref{pdistr}), $\rho$ in terms of Eq.~(\ref{rhodistr}), $g$ as given by Eq.~(\ref{gravfielddistr}), and finally
\begin{equation}
\tilde\xi (r,t)= \tilde\xi^{+}(r,t)\theta(r-R) + \tilde\xi^-\theta(R-r)\label{tildexidist},
\end{equation}
thus the equation governing the evolution of perturbations on a given volume element of the fluid is
\begin{equation}
\Delta\bigg\{\rho\frac{dv_r}{dt}+\frac{\partial P}{\partial r} + \rho g(r)
- \left[ \frac{2{\cal P}}{R} + \frac{\sigma}{2}(\dot{v}_r^+ + \dot{v}_r^-) \right]\delta(r-R)\bigg\}=0\label{masterpertdistrib}.
\end{equation}
Notice that we assumed that $\dot{v_r}$ is a distribution like $\tilde\xi$. Physically this must be taken because the phases are always ``localizable'' and do not mix. In other words, this constraint reflects the intuitive fact that the surface of discontinuity should be well-defined. The mathematical reason of why this is so shall be given below and it is related to the well-posedness of the problem.

When developed, taking into account the hydrostatic equilibrium equation [see Eq.~(\ref{hydrostaticeq})], it can be simplified to
\begin{eqnarray}
&&\frac{\partial}{\partial r}\left[\frac{\Gamma P}{r^2}\frac{\partial}{\partial r}(r^2\tilde\xi) \right] -\frac{4}{r}\frac{\partial P}{\partial r}\tilde\xi -\frac{d^2 \tilde\xi}{dt^2} \rho = \left[\frac{2\tilde{\xi}}{R^2}\left(2\eta^2\sigma - 3{\cal P}\right) \right.\nonumber\\ && \left. - \frac{\sigma}{2}\left(\frac{d^2\tilde{\xi}^+}{dt^2} + \frac{d^2\tilde{\xi}^-}{dt^2} \right) \right]\delta(r-R) \label{masterpertdistribexp}.
\end{eqnarray}
First note that coefficients multiplied by $\delta^2(r-R)$ or $\delta'(r-R)$ in Eq. (\ref{masterpertdistribexp}) must be all null.
Taking into account Eq. (\ref{tildexidist}), it means that
\begin{equation}
[\tilde\xi]^+_-=0.
\end{equation}
This automatically warrants $\ddot{\tilde{\xi}}$ as a distribution without Dirac deltas, as we have advanced previously. In order to obtain Eq.~(\ref{masterpertdistribexp}), we used the results that for a distribution $A(r)= A^+(r)\theta(r-R) + A^-(r)~\theta~(R-r)$,
\begin{eqnarray}
\Delta A &=& \delta A + \frac{\partial A}{\partial r} \tilde{\xi} - [A]^+_-\tilde{\xi}\delta(r-R)\label{Deltadistr},\\
\Delta \left(\frac{\partial A}{\partial r} \right) &=& \frac{\partial}{\partial r}(\Delta A) -\frac{\partial \tilde{\xi}}{\partial r}\frac{\partial A}{\partial r} \nonumber\\ &+& \frac{1}{2}\left[\frac{\partial \tilde{\xi}^+}{\partial r} + \frac{\partial \tilde{\xi}^-}{\partial r}\right][A(R)]^+_-\delta(r-R)\label{Deltaderivdistri}.
\end{eqnarray}
Besides the above mathematical properties, we have also made use of
\begin{equation}
\Delta\rho= -\frac{\rho}{r^2}\frac{\partial}{\partial r}(r^2\tilde\xi) + \frac{\sigma}{2}\left(\frac{\partial\tilde{\xi}^+}{\partial r} + \frac{\partial\tilde{\xi}^-}{\partial r} \right)\delta(r-R) \label{Deltarhodistr},
\end{equation}
which is a direct consequence of assuming that the total mass in each phase is constant, even in the presence of perturbations. This is just guaranteed if the surface of discontinuity is stable, a prime hypothesis for having a well-posed stability problem. We also have assumed that $P=P(\rho)$, which implies that $\Delta P^{\pm}= \Gamma^{\pm} P^{\pm} \Delta\rho^{\pm}/\rho^{\pm}$. In deriving Eq.~(\ref{masterpertdistribexp}), we further took into account Eq.~(\ref{surfacetension}). We finally stress that a simpler way to obtain Eq.~(\ref{masterpertdistribexp}) is to recall that $\Delta M =0$, which guarantees that $\Delta g = -2 g\tilde\xi /r$. The fact that $\Delta M = 0$ means that comoving observers with the fluid do not notice a mass change. The aforementioned result can also be directly showed by Eqs.~(\ref{gravfield}), (\ref{Deltaderivdistri}) and (\ref{Deltarhodistr}).

It can be seen that only solutions of the type $\tilde{\xi}^{\pm}(r,t)=e^{i\omega^{\pm} t}\xi^{\pm}(r)$ for Eq.~(\ref{masterpertdistribexp}) are just meaningful if
\begin{equation}
\omega^{+}=\omega^{-}\doteq \omega.
\end{equation}
This is the only way to eliminate the time dependence above in Eq.~(\ref{masterpertdistribexp}), and also to guarantee that the jump of $\tilde\xi$ be
null for any surface of discontinuity at any time. Therefore, we arrive at the important conclusion that even a stratified system where oscillatory
perturbations take place should be described by a sole set of frequencies. Each member of this set describes the eigenfrequency of whole system,
instead of one or another phase. Nevertheless, we recall that at the surface of discontinuity the frequencies are in principle not defined. Bearing
in mind the above conclusions, we have that, using Eq.~(\ref{eq:xi}),
\begin{equation}
\xi(r)=\xi^-(r)\theta(R-r) + \xi^+(r)\theta(r-R)\label{xidistr}
\end{equation}
and the boundary condition
\begin{equation}\label{contixi}
[\xi(R)]^+_-=0,\quad {\rm or}\quad \xi^+(R)=\xi^-(R)\doteq \xi(R),
\end{equation}
and therefore the only meaningful $\Gamma$ are given by
\begin{equation}
\Gamma(r)=\Gamma^-(r)\theta(R-r) + \Gamma^+(r)\theta(r-R)\label{Gammadistr}.
\end{equation}
Gathering the above equations on Eq.~(\ref{masterpertdistribexp}), we obtain
\begin{eqnarray}
&&\frac{d}{dr}\left[\frac{\Gamma P}{r^2}\frac{\partial}{\partial r}(r^2\xi) \right] -\frac{4}{r}\frac{d P}{d r}\xi +\omega^2 \rho\xi =\nonumber\\ && -\xi(R)\left[\frac{2}{R^2}\left(3{\cal P} -2\eta^2\sigma \right) - \omega^2\sigma  \right]\delta(r-R) \label{masterpertdistribsimpl}.
\end{eqnarray}
Ones sees from Eq.~(\ref{masterpertdistribsimpl}) that in the case where ${\cal P}$ and $\sigma$ are null, the classical expression, Eq.~(\ref{perteq}), is recovered.

Summing up, substituting Eqs.~(\ref{xidistr}) and (\ref{rhodistr}) into Eq.~(\ref{masterpertdistribsimpl}), one sees that the only way to satisfy such an equation is by imposing that
\begin{equation}
\frac{d}{dr}\left[\Gamma^{\pm} P^{\pm}\frac{1}{r^2}\frac{d}{dr}(r^2\xi^{\pm}) \right]-\frac{4}{r}\frac{dP^{\pm}}{dr}\xi^{\pm} +\omega^2\rho^{\pm}\xi^{\pm}=0\label{pertpm},
\end{equation}
\begin{equation}
\left[\Gamma P\frac{d}{dr}(r^2\xi) \right]^{+}_{-} + 2( 3{\cal P} - 2\eta^2\sigma - 2R[P(R)]^+_-){\xi(R)} =0 \label{descontderxi},
\end{equation}
and for completeness, condition (\ref{contixi}). Eq.~(\ref{pertpm}) is obtained here as a consequence of our distributional search for solutions to the radial Lagrangian displacements. This is exactly what one expects under physical arguments. Eqs.~(\ref{descontderxi}) and (\ref{contixi}) are our desired boundary conditions to be further taken into account
[besides Eq.~(\ref{boundarycont})] at the interface of any two phases.

For the case where $[P(R)]^+_-$ $=$ $[\Gamma(R)]^+_-$ $=$ $\sigma= {\cal P}=0$, we have that also the derivative of $\xi$ is continuous and therefore $\xi$ is a
differentiable function anywhere, as it should be since we are defining here a continuous system. Nevertheless, whenever the aforementioned conditions
do not take place, richer scenarios rise. Even in the case of a phase transition at constant pressure and negligible surface mass, the discontinuity
of $\Gamma$ and the existence of ${\cal P}$ generally render the derivative of $\xi$ discontinuous.

\section{A specific example: uniform density stars}\label{sec:3}

We would like to stress that the boundary condition we have derived previously is actually very restrictive. The reason for this is that the physically acceptable cases [solutions to Eqs. (\ref{pertpm}) associated with a surface of discontinuity at $r=R$] are only the ones that deliver enough arbitrary constants of integration for satisfying Eqs.~(\ref{xidistr}) and (\ref{descontderxi}). This should be taken into account together with the physical requirement of only admitting finite $\xi$ everywhere and that are null at the origin [see Eq.~(\ref{boundarycont})]. In the following we shall see a particular example where all of these aspects are evidenced.


Let us now investigate a star made out of two phases, each with a uniform mass-density. Let us assume also, for the sake of simplicity and exemplification, that the associated $\Gamma$ for each region is an arbitrary constant. As we will see, although this can be considered only as a first academic example, it already evidences some aspects stratified systems should have. For this case it is straightforward to solve Poisson's equation and the equation of hydrostatic equilibrium \citep[see, e.g.,][for further details]{1986bhwd.book.....S} and we have for $r<R$
\begin{equation}
P^-(r)=\frac{2\pi G\rho_-^2}{3}({\cal R}^2-r^2),\;\; {\cal R}\doteq \frac{3p_0^-}{2\pi G \rho_-^2}\label{phomminus},
\end{equation}
where $p_0^-$ is an arbitrary constant that corresponds to the pressure of system at the origin. For $r>R$, instead
\begin{equation}
P^+(r)=\frac{2\pi G\rho_+^2}{3}(R_s^2-r^2)\label{phomplus}.
\end{equation}
The constant mass-density in the inner and outer regions has been defined as $\rho^-$ and $\rho^+$, respectively. The pressure at the origin $p_0^-$ could always be chosen such that it matches the pressure at the base of the outer phase, and as it can be seen from Eq.~(\ref{phomplus}), we have also introduced the condition of having a null pressure at the star's surface. Substituting Eqs.~(\ref{phomminus}) and (\ref{phomplus}) into Eq.~(\ref{pertpm}), we are led to
\begin{equation}
(1-x^2_{\pm})\frac{d^2\xi^{\pm}}{dx_{\pm}^2}+\left(\frac{2}{x_{\pm}}-4x_{\pm} \right)\frac{d\xi^{\pm}}{dx_{\pm}}+\left(A_{\pm}-\frac{2}{x^2_{\pm}} \right)\xi^{\pm}=0\label{pertpmhom},
\end{equation}
where we assumed that $x_{+}\doteq r/R_s$, $x_{-}\doteq r/{\cal R}$, and
\begin{equation}
A_{\pm}\doteq \frac{3\omega^2_{\pm}}{2\pi G \rho^{\pm} \Gamma^{\pm}}+\frac{8}{\Gamma^{\pm}}-2\label{Apm}.
\end{equation}
We now solve Eq.~(\ref{pertpmhom}) by the method of Frobenius. For a sake of simplicity, we shall drop the $\pm$ notation.
We therefore assume solutions of the form
\begin{equation}
\xi= \sum_{n=0}^{\infty}a_n x^{n+s}\label{pseriesxihom},
\end{equation}
where $a_n$ and $s$ are arbitrary constants to be fixed by primarily demanding that the first condition of Eq.~(\ref{boundarycont}) is satisfied, as well as $\xi(x)$ is always finite. By substituting Eq.~(\ref{pseriesxihom}) into Eq.~(\ref{pertpmhom}), it can be checked that the solutions to $s$ are either $s=1$ or $s=-2$. The associated recurrence relation obtained generally is
\begin{equation}
a_{m+2}=\frac{(m+s)(m+s+3)-A}{(m+s+2)(m+s+3)-2} a_m,\label{am2am}
\end{equation}
with $m=0,2,4...,$ and $a_1=a_3=a_5=...=0$. Let us analyze first the inner region. It is clear in this case that the associated $a_0$ for $s=-2$ must be null, as a consequence of one of our boundary conditions. From Eq.~(\ref{am2am}), one clearly sees that the power series given by Eq.~(\ref{pseriesxihom}) does not converge. Therefore, in order to satisfy the finiteness anywhere of $\xi$, we have to impose that the series be truncated somewhere, rendering it actually a polynomial. Hence
\begin{equation}
A_{m,s=1}^-=(m+1)(m+4)\label{Amm}.
\end{equation}
From Eq.~(\ref{Apm}), one sees that just discrete frequencies [given by Eq.~(\ref{Amm})] are possible to this region. From Eqs.~(\ref{Apm}) and (\ref{Amm}), for having the frequency of the fundamental mode ($m=0$) positive, one should have $\Gamma^-\geq 4/3$. Summing up, the physically relevant solution to this case just leaves out {an} arbitrary constant of integration, as required due to the scaling law present to $\xi$ from Eq.~(\ref{pertpm}).

Let us now analyze the outer region. This is the most physically interesting region since the problems at the star's center are absent and therefore in principle one could even have two linearly independent solutions to $\xi$. Due to the finiteness of $\xi$ in this region, the outer counterpart of Eq.~(\ref{Amm}) must again take place. Nevertheless, for $s=-2$, one should also impose
\begin{equation}
A^+_{m,s=-2}=(m-2)(m+1)\label{Amp1}.
\end{equation}
From Eq.~(\ref{Apm}), one sees from this case that its associated fundamental mode ($m=0$) is unstable.
This means in principle that this solution to the outer region should be excluded, leaving out just the one from the case $s=1$, where we should consider
\begin{equation}
A^+_{m,s=1}= (m+1)(m+4)\label{Amp2}.
\end{equation}

Notwithstanding, our previous analysis exhibits clear problems: there are not enough arbitrary constants to fix Eqs.~(\ref{contixi}) and (\ref{descontderxi})  and the eigenfrequencies in each region are different. However, we shall show that the condition of having a same eigenfrequency for the whole system, as required by our formalism, addresses all the problems. Obviously the stable eigenfrequencies of the star are only related to the solution $s=1$. However, they could rise here either from aspects of the inner or the outer phases of the star. Let us see how this conclusion ensues. Assume initially that the
only possible {$\omega$'s are given by Eq.~(\ref{Amm}), associated with the modes $m^-_{s=1}$}. So, for having finite $\xi^+$ related to $s=1$,
one must impose that there exists a {$m^+_{s=1}$ to the outer phase} such that the numerator of the associated recurrence relation be null. It can be shown that this is just the case if
\begin{equation}
m_{s=1}^+=\frac{-5+\sqrt{9+ 4A_{s=1}^{+}(m_{s=1}^-)}}{2}\label{ms1}.
\end{equation}
Therefore, Eq.~(\ref{ms1}) demands that
\begin{equation}
9+ 4A_{s=1}^{+}=(2p+1)^2,\quad p\geq 2,\quad  p \in N \label{Gammapcond}.
\end{equation}
For the case $s=-2$ to $\xi^+$, it can be shown that the condition for the existence of a $m^+_{s=-2}$ related to {$a_{{m}_{s=-2}^+}=0$} is
exactly given by Eq.~(\ref{Gammapcond}). The mode itself is
\begin{equation}
m_{s=-2}^+ = \frac{1+\sqrt{9+ 4A_{s=1}^{+}(m_{s=1}^-)}}{2}\label{msm2}.
\end{equation}

Summarizing: if Eq.~(\ref{Gammapcond}) is satisfied for any natural $p\geq 2$, there always exist modes, characterized by Eqs.~(\ref{ms1}) and (\ref{msm2}), which guarantee the finiteness of $\xi^{+}$ as a linear combination of solutions for $s=1$ and $s=-2$, associated with a given eigenfrequency $\omega_{m^-_{s=1}}$ that just takes into account aspects of the inner phase of the system. In this case, one is able to come out with two arbitrary constants of integration, that would then guarantee that the additional boundary conditions raised by the stratification, Eqs.~(\ref{contixi}) and (\ref{descontderxi}), be satisfied. It is immediate to see that a similar reasoning as above ensues if one chooses now $\omega_{m^+_{s=1}}$ as coming from aspects of the outer region, given now by Eqs.~(\ref{Apm}) and (\ref{Amp2}). For this case, we will find now a $m^-_{s=1}$ and a ${m^+_{s=-2}}$ associated with $\omega_{m^+_{s=1}}$, as given by Eqs.~(\ref{ms1}) and (\ref{msm2}), with the condition given by Eq.~(\ref{Gammapcond}), replacing $A_{s=1}^{+}(m_{s=1}^-)$ by $A_{s=1}^{-}(m_{s=1}^+)$. Since $\Gamma^{\pm}$ and $\rho^{\pm}$ are given quantities, one sees that the only possible eigenfrequencies to system should satisfy $9+ 4A_{s=1}^{\pm}=(2p+1)^2$. This constraint is uniquely imposed due to the extra boundary conditions to the problem and is very restrictive. We have just shown a simple example where some of the aspects imprinted by stratification raise. Whenever there are two arbitrary solutions to $\xi$ in a given phase, it will be always possible to satisfy the constraints (\ref{contixi}) and (\ref{descontderxi}).

\section{Systems with an electromagnetic structure}\label{sec:5}

Now we attempt to give a further step in our classical generalization, by endowing the phases (as well as the surface of discontinuity) with an electromagnetic structure. Just for clarity, let us work with a system that exhibits {just} an electric field. The first point to be taken into account is the additional electric force present in the system. This would have the same structure as the gravitational force and therefore its generalization is straightforward. Now one should define also a distributional solution to the charge density. The surface force associated with the surface tension should have the same form as previously, but now should also take into account the present electric aspects. The pressure in this case would also change due to the presence of the electric field and its jump over a surface of discontinuity could still be kept free.

From the (distributional) Maxwell equations in the spherically symmetric case one has that
\begin{equation}
E(r)= \frac{Q(r)}{r^2},\quad Q(r)\doteq 4\pi \int_0^r \rho_c(\bar r)\bar{r}^2d\bar r \label{electricf},
\end{equation}
where $\rho_c(r)$ is the charge density at $r$. The associated ``force density'' is $d\vec{F}_{el}/dv= \rho_c E(r)\hat r$. Therefore, the equation
of hydrostatic equilibrium now reads
\begin{equation}
\frac{d P}{dr} + \rho(r) g(r) - \rho_c(r)E(r) - \frac{2{\cal P}_Q}{R}\delta(r-R)=0 \label{hydrostaticeqelec}.
\end{equation}

Therefore, like the gravitational field, the electric field also presents a jump at any surface of discontinuity (at $r=R$) endowed with surface charges. We write the charge-density as
\begin{equation}
\rho_c(r)= \rho^-_c(r)\theta(R-r) + \rho^+_c(r)\theta(r-R) + \sigma_c\delta(r-R)\label{chargedensdistr},
\end{equation}
while the distributional electric field is
\begin{equation}
E(r)= E^-(r)\theta(R-r) + E^+(r)\theta(r-R),\; [E(R)]^+_- = 4\pi \sigma_c\label{electricfdistr}.
\end{equation}
By substituting now Eqs.~(\ref{pdistr}), (\ref{electricf}), (\ref{chargedensdistr}) and (\ref{electricfdistr}) into Eq.~(\ref{hydrostaticeqelec}),
we have that the surface tension at equilibrium should read
\begin{equation}\label{surfacetensionef}
{\cal P}_Q = \frac{R}{2}[P(R)]^+_- + \frac{G}{16\pi R^3}[M^2(R)]^+_-  - \frac{1}{16\pi R^3}[Q^2(R)]^+_- .
\end{equation}
Notice that the existence of a surface mass would lead to $[M^2(R)]^+_->0$, while $[Q^2(R)]^+_-$ could in principle be any. The appearance of the last term
in Eq.~(\ref{surfacetensionef}) is consistent with the expected and long ago known contribution of electric double-layers on the surface tension and
surface energy of metals, as recalled by \citet{frenkel1917} in his seminal work. The existence of such surface electric fields is well-known in
material sciences and it has been determined experimentally from the photoelectric phenomenon by measuring the amount of work done by electrons to escape
from the metal's surface. There is a vast literature on the role of electric double-layers on surface phenomena in metals and contact surfaces, and we
refer the reader for instance to \citet{huang1949,surfaceforces2011}, and references therein, for further details on this subject.

Since in the presence of an electric field the hydrostatic equilibrium equation and the surface tension changes, it can be checked that Eq.~(\ref{masterpertdistribsimpl}) keeps the same functional form. In drawing this conclusion, it was also assumed that the total charge of the system is a constant. This also means that $\Delta Q=0$. One also sees immediately that the main results concerning the stability of the stratified charged case are totally analogous to the neutral one, obtained by simply making the replacement
${\cal P}\rightarrow {\cal P}_{Q}$.

\section{Stratified systems in general relativity}\label{sec:6}

Now we generalize the analysis of stratified systems to general relativity. From the classical analysis, we have learned that surface
quantities must also be inserted into the generalized equation of hydrostatic equilibrium. Therefore, in a certain sense, we must find the proper generalization of the
surface forces in general relativity. This will not be difficult bearing in mind the thin shell formalism, as we shall see in the sequel. Such a
formalism states that in order to search for distributional solutions to general relativity, one has to consider an energy-momentum tensor at a
surface of discontinuity, that we shall name $\Sigma$. It is precisely this surface content that leads to the jump of quantities that are related to physical observables, such
as the extrinsic curvature. We now outline the formalism succinctly. Let us work just in the spherically symmetric case, where $\Sigma$ is defined as $\Phi= r-R(\tau)=0$, with $\tau$ the the proper time of an observer on the aforesaid hypersurface.
Assume that the metrics in the regions above and below $\Sigma$
(w.r.t. to the normal vector to it), described by  the coordinate systems $x^{\mu}_{\pm} \doteq (t_{\pm}, r_{\pm}, \theta_{\pm}, \varphi_{\pm})$,
respectively, are given by
\begin{equation}
ds^2_{\pm}=e^{2\alpha_{\pm}(r_{\pm})}dt^2_{\pm}-e^{2\beta_{\pm}(r_{\pm})}dr^2_{\pm}-r^2_{\pm}d\Omega^2_{\pm}\label{linelsphpm},
\end{equation}
where
\begin{equation}
d\Omega^2_{\pm}= d\theta^2_{\pm} + \sin^2\theta_{\pm}d\varphi^2_{\pm}\label{solangle}.
\end{equation}
Assume that the (three dimensional) hypersurface $\Sigma$ be described by the (intrinsic) coordinates $y^a \doteq (\tau, \theta, \varphi)$ such
that at the hypersurface $t_{\pm} =t_{\pm} (\tau)$, $\theta_{\pm}=\theta$ and $\varphi_{\pm}=\varphi$, besides obviously $r_{\pm} = R(\tau)$.
In order to render the procedure consistent, one has to impose primarily that the intrinsic metric to $\Sigma$ is unique. This fixes the
coordinate transformations $x_{\pm}^{\mu} = x_{\pm}^{\mu} (y^a)$. This is the generalization of the continuity of the gravitational potential
across a surface {harboring} surface degrees of freedom. Now, if the jump of the extrinsic curvature is non-null, it is automatically guaranteed
the existence of a surface energy-momentum tensor \citep{2004reto.book.....P} that in the spherically symmetric case can always be cast as
$S^{a}{}_{b}= {\rm diag}(\sigma, -{\cal P}, -{\cal P})$, with \citep[see, e.g.,][]{2005CQGra..22.4869L}
\begin{equation}
\sigma = -\frac{1}{4\pi R}\left[\sqrt{e^{-2\beta}+\dot R^2}\right]^+_-\label{enerdens},
\end{equation}
\begin{equation}
{\cal P} = -\frac{\sigma}{2} + \frac{1}{8\pi R}\left[\frac{R\alpha'(e^{-2\beta}+\dot R^2)+\ddot R R + \beta' R\dot R^2}{\sqrt{e^{-2\beta}+\dot R^2}}\right]^+_-\label{surfdens},
\end{equation}
where generically $A'\doteq \partial A/\partial r$ and $\dot{A}\doteq dA/d\tau$. Finally, the discontinuity of the extrinsic curvature is the generalization of the discontinuity of the gravitational field {across a surface with nontrivial degrees of freedom}. The case of interest to be analyzed here is the static and stable (upon radial displacements of $\Sigma$) one $\dot{R}=\ddot{R}=0$, i.e. an equilibrium point.

Let us see now how the generalization of the surface forces appear in this formalism. First of all, we know that
\begin{equation}
T_{\mu\nu}= T_{\mu\nu}^{+}\theta(r-R) +T_{\mu\nu}^{-}\theta(R-r) + e^{a}_{\mu}e^{b}_{\nu}S_{ab} \delta(r-R)\label{tmunudistr},
\end{equation}
where $e^{a}_{\mu}\doteq \partial y^a/\partial x^{\mu}$ and $S_{ab}= h_{ac}S^{c}_b$, $h_{ab}\doteq e^{\mu}_a e^{\nu}_b g_{\mu\nu}$. Notice that $e^{a}_{\mu}$ is
itself defined as a distribution. For $h_{ab}$, it does not matter the side of $\Sigma$ one takes to evaluate it, since it must be unique.
Let us constrain ourselves first to the case of perfect fluids (locally neutral) on each side of $\Sigma$. One sees from Eq.~(\ref{tmunudistr})
and the coordinate transformations at $\Sigma$ that
\begin{equation}
T^0_0\doteq \rho= \rho^+ \theta(r-R) +  \rho^- \theta(R-r) + \sigma \delta(r-R) \label{t00distr},
\end{equation}
\begin{equation}
T^1_1 \doteq  -P= -P^+ \theta(r-R) -P^- \theta(R-r)\label{t11distr}
\end{equation}
and
\begin{equation}
T^2_2= T^3_3\doteq -P_t= -P -  {\cal P} \delta(r-R)\label{t22t33distr}.
\end{equation}
From Eq.~(\ref{t11distr}), we notice that there are not associated surface stresses. {This is exactly what we advanced in the classical case with heuristic arguments and obtained here as a general consequence of distributional solutions to general relativity.} Let us search formally for solutions to Einstein's equations with the energy-momentum given by Eqs.~(\ref{t00distr})--(\ref{t22t33distr}) with the ansatz
\begin{equation}
ds^2= e^{2\alpha}dt^2-e^{2\beta}dr^2-r^2(d\theta^2+ \sin^2\theta d\varphi^2)\label{metricgen}.
\end{equation}
The distributional nature of Eq.~(\ref{metricgen}) will be evidenced by Eq.~(\ref{tmunudistr}). As the solution, we have
\begin{equation}
e^{-2\beta}= 1-\frac{2m(r)}{r},\;\;m(r)\doteq 4\pi \int^{r}_0 \rho(\bar{r})\bar{r}^2 d \bar{r}\label{g11distr}.
\end{equation}
Notice from the above equation that $\rho$ is given by Eq.~(\ref{t00distr}) and therefore $e^{-2\beta}$ is a distribution generally discontinuous at $R$.
For $\alpha$ we have, though
\begin{equation}
\alpha'= \frac{e^{2\beta}}{r^2}[4\pi P r^3 + m(r)]\label{alphaprime},
\end{equation}
where $P$ is given by Eq.~(\ref{t11distr}). From Eqs.~(\ref{g11distr}) and (\ref{alphaprime}), we see that $\alpha'$ is a distribution with no Dirac delta functions terms. Therefore, it implies that $[\alpha]^+_-=0$. In other words, the function $\alpha$ is {generally} continuous though not differentiable at $R$.
Nevertheless, from the conservation law of the energy-momentum tensor given by Eq.~(\ref{tmunudistr}), we also have
\begin{equation}
\alpha'_{\pm}(\rho_{\pm} + P_{\pm})= -P'_{\pm}\label{alphaprimepressurepm}.
\end{equation}
We notice that $T^{\mu\nu}{}_{;\nu}=0$, taking into account Eq.~(\ref{tmunudistr}), would give us in principle terms dependent upon Heaviside functions, Dirac delta functions and their derivatives. The terms associated with the Heaviside functions are null due to the validity of Einstein's equations on each side of $\Sigma$. The nullity of the remaining terms is associated with identities that the surface energy-momentum tensor has to satisfy \citep[see, e.g.,][]{1996JMP....37.5672M}. It is not difficult to show that such identities are automatically satisfied when one takes into account Eqs.~(\ref{enerdens}) and (\ref{surfdens}) \citep[see, e.g.,][]{2005CQGra..22.4869L}. This shows that the thin-shell formalism is consistent and the surface terms must indeed be taken as the aforesaid equations.

In order to put Eq.~(\ref{alphaprimepressurepm}) in the form that would allow us to consider Eqs.~(\ref{t00distr}) and (\ref{t11distr}),
we mandatorily should add surface terms. The correct way of doing it is
%
%
%
%
\begin{eqnarray}
&&\frac{d P}{dr}= -{(\rho + P)}\alpha' + \frac{2 {\cal P}}{R}\delta(r-R) + \bigg\{[P(R)]^+_- + \nonumber \\ &&
\frac{\sigma}{2}[\alpha'_+(R) + \alpha'_-(R)] + \frac{\sigma}{R} - \frac{[\alpha' e^{-\beta}]^+_-}{4\pi R} \bigg\}\delta(r-R)\label{hydrostaticeqgr}.
\end{eqnarray}

Now we show that in the classical limit, Eq.~(\ref{hydrostaticeqgr}) reduces exactly to Eq.~(\ref{hydrostaticeq}) and thus it is
its proper generalization. First of all, notice that in such a limit, $\alpha = \phi (r)$, $\phi(r)$ the gravitational potential.
Besides, it can be shown that in such a case
\begin{equation}
\sigma= \frac{1}{4\pi R^2}[M(R)]^+_-\label{surfdensclasslim},
\end{equation}
where the above quantities are in cgs units. In Eq.~(\ref{surfdensclasslim}) one recognizes the jump of the gravitational field $g(r)=\phi'= GM(r)/r^2$ at $r=R$, as exactly given by Eq.~(\ref{gravfieldjump}). Therefore,

\begin{eqnarray}
\frac{\sigma}{2}\{\alpha'_+(R)+\alpha'_-(R)\}\simeq \frac{G [M^2(R)]^+_-}{8\pi R^4}\\\label{sigmaalphaprime}.
\end{eqnarray}
Substituting Eq.~(\ref{sigmaalphaprime}) into Eq.~(\ref{hydrostaticeqgr}), we see that the term inside the curly brackets of the latter equation is null [see Eqs.~(\ref{surfdens}) and (\ref{surfacetension})]. Hence, the remaining term in front of the delta function is exactly $2{\cal P}/R$ as we
{already advanced and expected} [see Eq.~(\ref{hydrostaticeq})].

Now we are in a position to talk about perturbations in the general relativistic scenario. When they take place, metric and fluid quantities change at a given spacetime point from their static counterparts. It is customary to assume that such departures are small, what allows us to work perturbatively. The primary task is to find such changes from the system of equations coming from relativistic hydrodynamics and general relativity. Nevertheless, these solutions are already very well-known \citep{1973grav.book.....M}. Our ultimate task is simply to generalize them to the distributional case.

The equation governing the evolution of the fluid displacements on each side of $\Sigma$ is the general relativistic Euler equation, related to the orthogonal projection of $T^{\mu\nu}{}_{;\nu}$ (perfect fluids) onto $u^{\mu}$, i.e.
\begin{equation}
(\rho + P)u^{\nu}{}_{;\,\mu}u^{\mu}\doteq (\rho + P)a^{\nu}= (g^{\mu\nu}-u^{\mu}u^{\nu})P_{,\mu}\label{releulereq},
\end{equation}
where the labels $``\pm''$ for each term in the above equation were omitted just not to overload the notation.

In the hydrostatic case we have only that $u^{t}_{\pm}= e^{-\alpha_0^{\pm}}$. When perturbations are present \citep{1973grav.book.....M}
\begin{equation}
u^{t}_{\pm}= e^{-\alpha^{\pm}} = e^{-\alpha_0^{\pm}}(1-\delta\alpha^{\pm}) \label{ut},
\end{equation}
where $\delta\alpha$ is the change of the static solution $\alpha_0$ in the presence of perturbations at a given spacetime point. For the $u^r_{\pm}$ component,
using the normalization condition $u^{\mu}_{\pm}u_{\mu}^{\pm}=1$, one shows that \citep{1973grav.book.....M}
\begin{equation}
u^{r}_{\pm}= e^{-\alpha_0^{\pm}}\dot{\tilde\xi}_{\pm}\label{ur},
\end{equation}
with $\dot{\tilde\xi}^{\pm}\doteq \partial \tilde\xi^{\pm}/\partial t^{\pm}$. Just for completeness, $u^{\theta}_{\pm}=u^{\varphi}_{\pm}=0$. For the components of $u^{\mu}_{\pm}$ given by Eqs.~(\ref{ut})
and (\ref{ur}), the left-hand side of Eq.~(\ref{releulereq}) gives as the only nontrivial component $a^{r}_{\pm}=-e^{-2\beta^{\pm}}a_{r}^{\pm}$, with
\begin{equation}
-a_{r}^{\pm}= \alpha'_{0\pm} + \delta\alpha'_{\pm} + e^{2(\beta_0^{\pm}-\alpha^{\pm}_0)}\ddot{\tilde\xi}_{\pm}\label{ar}.
\end{equation}
and the associated equation of motion
\begin{equation}
(\rho^{\pm} + P^{\pm})(-a_r^{\pm})= -\frac{\partial P^{\pm}}{\partial r_{\pm}}\label{arfinal}.
\end{equation}
From Eq.~(\ref{ar}), Eq.~(\ref{arfinal}) can be cast as
\begin{equation}
(\rho^{\pm}_0+P^{\pm}_0)e^{2(\beta_0^{\pm}-\alpha_0^{\pm})} \ddot{\tilde\xi}_{\pm}= -\frac{\partial P^{\pm}}{\partial r^{\pm}}- (\rho^{\pm}+P^{\pm})\alpha'_{\pm}\label{xiddot}.
\end{equation}
Therefore, in terms of distributions Eq.~(\ref{xiddot}) reads
\begin{widetext}
\begin{eqnarray}
(\rho_0+P_0)e^{2(\beta_0-\alpha_0)} \ddot{\tilde\xi}= &-&\frac{\partial P}{\partial r}- (\rho+p)\alpha' +  \frac{2{\cal P}}{R} \delta(r-R) + \bigg[ \frac{\sigma}{R} - \frac{[\alpha' e^{-\beta}]^+_-}{4\pi R} +  [P(R)]^+_- \nonumber \\
&+& \frac{\sigma}{2}\left\{e^{2(\beta_0^{+}-\alpha_0^{+})} \ddot{\tilde\xi}_{+} + e^{2(\beta_0^{-}-\alpha_0^{-})} \ddot{\tilde\xi}_{-} + \alpha'_+ + \alpha'_-\right\}\bigg]\delta(r-R)\label{xiddotdistr},
\end{eqnarray}
\end{widetext}
where $\rho$ and $P$ are given by Eqs.~(\ref{t00distr}) and (\ref{t11distr}), respectively, and now
\begin{equation}
\tilde\xi(r,t)= \tilde\xi^+(r^+,t^+)\theta(r^+-R) + \tilde\xi^-(r^-,t^-)\theta(R-r^-)\label{xirtdistri}.
\end{equation}
Notice that in Eq.~(\ref{xiddotdistr}) we are considering jumps and symmetrizations of quantities defined in the presence of perturbations. As we stated previously, such perturbations change slightly the value of the physical quantities with respect to their hydrostatic values. The square brackets term of Eq.~(\ref{xiddotdistr}) is the proper generalization of the curly brackets term in Eq.~(\ref{masterpertdistrib}). Naturally the reasoning for the eigenfrequencies of $\tilde\xi$ in the general relativistic is the same as in the classical case. The same can be said about the continuity, though not differentiability, of $\tilde\xi$ at any surface of discontinuity.

Now, in order to have the proper generalization of Eq.~(\ref{masterpertdistribsimpl}), we should evaluate Eq.~(\ref{xiddotdistr}) at $r+\tilde\xi$ and then subtract it from its evaluation at $r$ concerning the static solution. In order to do it properly, one should take into account the general results for the case of Lagrangian displacements coming from the standard procedure \citep[see, e.g.,][]{1973grav.book.....M}, but now in the sense of distributions, by recalling that $\Delta A\doteq A(r+\tilde\xi,t)-A_0(r,t)$, where $A_0$ concerns the quantity $A$ at equilibrium.

It is not difficult to see that we have the following results in the general relativistic distributional case \citep[see, e.g.,][for the
treatment of a continuous system]{1973grav.book.....M}:
\begin{eqnarray}
\Delta \beta &=& -\alpha'_{0}\tilde\xi\label{dbeta},\\
\Delta\alpha' &=& 4\pi r e^{2\beta_{0}}\left[ \delta P + 2\delta \beta P_0 \right] + \frac{e^{2\beta_{0}}}{r}\delta\beta  +
\alpha'_{0}\tilde\xi \label{dalphaprime},\\
\Delta P &=& -\gamma P_0 \left[ \frac{e^{-\beta_{0}}\left(r^2e^{\beta_{0}}\tilde\xi\right)'}{r^2} + \delta\beta \right]\label{dp},\\
\Delta\rho &=& -(\rho_0 + P_0) \left[\frac{1}{r^2}\frac{\partial}{\partial r}(r^2\tilde\xi) \right] +
\bigg(\frac{\sigma_0}{2}\bigg\{\frac{\partial \tilde{\xi}^+}{\partial r} + \frac{\partial \tilde{\xi}^-}{\partial r} \bigg\} \nonumber\\ &&
+ \Delta\sigma + \frac{2\sigma_0 \tilde{\xi}}{R} + [P_0]^+_-\tilde{\xi} \bigg)\delta(r-R) -\frac{d P_0}{d r}\tilde{\xi}\label{drho},\\
\gamma &\doteq& \frac{\rho_0+ P_0}{P_0}\left(\frac{\partial P}{\partial \rho}\right)_{s={\rm const}}\label{gamma},\\
\Delta \sigma &=& -\frac{2\sigma_0 \tilde{\xi}}{R} - \tilde{\xi}\left([e^{\beta_0}P_0]^+_- + \frac{[\cosh\beta_0]^+_-}{4\pi R^2} \right)\label{dsigmagr}.
\end{eqnarray}
We stress that Eq.~(\ref{dp}) assumes adiabatic processes, in which one {considers} $P=P(\rho)$ and we have made use of $[\tilde\xi]^+_-=0$ for the above equations.

We shall seek for solutions to the perturbations as $\tilde\xi= e^{i\omega t} \xi(r)$ with $\omega$ the same for all the phases the system may have.
We just need to worry about the Dirac delta function term, since it gives us the desired boundary condition valid for the separation of each two phases.
The terms in front of the Heaviside functions by default shall be the ones found in continuous media. It is not hard to see that the surface terms at
the end should satisfy the condition
\begin{equation}
\frac{\Delta\sigma}{R}(2\eta^2+1) - \frac{\Delta[\alpha'e^{\beta}]^+_-}{4\pi R}=0\label{deltatermgr},
\end{equation}
where we have used Eq.~(\ref{Deltaderivdistri}). When the last term on the left-hand side of above equation is expanded, by using Eqs.~(\ref{alphaprime})
and (\ref{dbeta}), Eq.~(\ref{deltatermgr}) can be further simplified to
\begin{equation}
\frac{2\eta^2\Delta\sigma}{R} - \Delta\left[\frac{[\cosh\beta]^+_-}{4\pi R^2}\right] = \left[ \frac{2{\cal P}}{R^2} - [Pe^{\beta}\alpha']^+_-\right]\tilde\xi + [\Delta P e^{\beta_0}]^+_- \label{deltatermgrsimp}.
\end{equation}
One sees from Eq.~(\ref{deltatermgrsimp}) that Eq.~(\ref{descontderxi}) is recovered in the classical limit by recalling that $\beta=M(r)/r$, which implies that the last term on the left-hand side of the above equation is $4\sigma (g^+ + g^-)/(2R)$. Besides, in this limit we take $P\rightarrow 0$ and  $e^{\beta}\rightarrow 1$ for the remaining terms.

The case where electromagnetic interactions are also present is also of interest since its associated energy-momentum tensor is anisotropic. This naturally influences the equation of hydrostatic equilibrium, since it now becomes
\begin{equation}
\alpha'(P + \rho) = -P'-\frac{2}{r}(P_t-P)\label{hydrostaticeqanis},
\end{equation}
where $P$, $P_t$ and $\rho$ are the resultant radial pressure, tangential pressure and energy density of the fluid, respectively. For the electromagnetic fields, clearly $(P_t-P)$ is solely related to them. Due to the aforementioned aspects, the latter should also influence the dynamics of the radial perturbations, as we shall show in the next section.

\section{Electromagnetic interactions in stratified systems within GR}\label{sec:7}

We consider now the inclusion of electromagnetic interactions within the scope of stratified systems in general relativity. An important comment at this level is in order. Since we are dealing with electromagnetic fields in stars, it would be more reasonable to assume the Maxwell equations in material media. Nevertheless, since the knowledge of the structures constituting the stars is not yet precise, it is difficult to assess their realistic dielectric properties. Since working with Maxwell equations in the absence of material media gives us upper limits to the fields under normal circumstances, this seems to be a good first tool to evaluate the relevance and effects of electromagnetism in stars. We shall adopt for the time being to follow this approach. The energy-momentum tensor of each layer of the system we are now interested should also have the electromagnetic one \footnote{We restrict our analyses to the Maxwell Lagrangian, $-F^{\mu\nu}F_{\mu\nu}/4\doteq -F/4$.}, i.e.
\begin{equation}
4\pi T_{\mu\nu}^{(em)}= -F_{\mu\alpha}F_{\nu\beta}g^{\alpha\beta} + g_{\mu\nu}\frac{F^{\alpha\beta}F_{\alpha\beta}}{4}\label{tmunumax},
\end{equation}
where we defined $F_{\mu\nu}\doteq \partial_{\mu}A_{\nu}-  \partial_{\nu}A_{\mu}$. Solving Einstein-Maxwell equations on each layer of a stratified system leads us to the following equilibrium condition \citep{1971PhRvD...4.2185B}
\begin{equation}
\frac{\partial P}{\partial r} = \frac{Q(r)Q'(r)}{4\pi r^4} - \alpha'_Q (\rho + P) \label{eqhydchargedTOV},
\end{equation}
where
\begin{equation}
Q(r)\doteq \int_{0}^r 4\pi r^2 \rho_{c}e^{\beta_Q}dr,\;\; E(r)= e^{\alpha_Q+\beta_Q}\frac{Q(r)}{r^2}\label{Qdef},
\end{equation}
\begin{equation}
e^{-\beta_Q}= 1-\frac{2m_Q(r)}{r}+ \frac{Q^2(r)}{r^2}\label{ebetaq}
\end{equation}
\begin{equation}
\alpha'_Q = \frac{e^{2\beta_Q}}{r^2}\left[4\pi r^3 P + m_Q(r) - \frac{Q^2(r)}{r} \right]\label{alphaprimeQ}
\end{equation}
and
\begin{equation}
m_Q(r)\doteq \int^{r}_0 4 \pi r^2 \rho dr + \frac{Q^2}{2r} + \frac{1}{2}\int^{r}_0 \frac{Q^2}{r^2}dr \label{mQ}.
\end{equation}
Eqs.~(\ref{Qdef}), (\ref{ebetaq}), (\ref{alphaprimeQ}) and (\ref{mQ}) are the charge, the radial and time components of the metric and the energy of the system up to a radial coordinate $r$, respectively. Besides, in Eq.~(\ref{Qdef}), we also exhibited the electric field $E(r)$ in the context of general relativity, obtained by means of the definition $F_{tr}\doteq E(r)= -\partial_r A_0$. We stress that $\rho_{c}$ is the physical charge density of the system,
defined in terms of the four-current by $j^{\mu}\doteq e^{-\alpha_Q}u^{\mu}$, where $u^{\mu}$ is the four-velocity of the fluid with respect to the coordinate system $(t,r,\theta,\varphi)$ \citep[see][for further details]{1975ctf..book.....L}.

From Eq.~(\ref{eqhydchargedTOV}) one sees that in the scope of general relativity the effect of the charge is not merely to counterbalance the gravitational pull. For certain cases, it could even contribute to it. The reason for that is due to the contribution of the electromagnetic energy to the final mass of the system, as clearly given by Eq.~(\ref{mQ}). Notice from Eq.~(\ref{mQ}) that we have assumed that the mass at the origin is null, in order to avoid singularities there. More generically, one could assume point or surface mass contributions in Eq.~(\ref{mQ}) by conveniently adding Dirac delta functions in $\rho$. Finally, we stress that Eq.~(\ref{Qdef}) can indeed be seen as the generalization of the charge in general relativity, since it takes into account the nontrivial contribution coming from the spacetime warp due to its energy-momentum content.

Notice that the classical limit to Eq.~(\ref{eqhydchargedTOV}) can be shown to coincide with Eq.~(\ref{hydrostaticeqelec}), by recalling that $E_{clas}(r)= Q_{clas}(r)/r^2$ and, from Eq.~(\ref{Qdef}), $Q'_{clas}(r)=4\pi r^2 \rho_{c}$. Besides, we recall that when brought to cgs units, the term $Q^2/r$ (here in geometric units) becomes $Q^2/(c^2r)$, which is null in the classic non-relativistic limit, as well as any pressure term on the right-hand side of the aforementioned equation.

Now, consider the analysis of a charged system constituted of two parts, connected by a surface of discontinuity (at $r^{\pm}=R$) which hosts surface degrees of freedom, such as an energy density, charge density and a surface tension. {Its generalization to an arbitrary number of layers is immediate since each surface of discontinuity is only split by two phases.} The proper description of the charge density in this case would be given by the generalization of Eq.~(\ref{chargedensdistr}). Therefore, one would have at equilibrium that
\begin{equation}
Q(r)= Q^-(r^-)\theta(R-r^-) + Q^+(r^+)\theta(r^+ - R) \label{Qdistrib}.
\end{equation}
and for $Q'(r)$, a Dirac delta shall rise, due to $\rho_{c}$.

Let us define the distribution
\begin{equation}
\bar{\rho}_c \doteq \rho_c e^{\beta_Q}\doteq \bar{\rho}_c^+\theta(r-R) + \bar{\rho}_c^-\theta(R-r) + \bar{\sigma}_c\delta(r-R)\label{Euclidiancden},
\end{equation}
where $\bar{\rho}_c^{\pm}$= $\bar{\rho}_c^{\pm}(r^{\pm})$. From the above definition, we have that $Q'=4\pi r^2 \bar{\rho}_c$. It implies that the total charge is the same as the one associated with $\bar{\rho}_c$ in an Euclidean space. Therefore, all classical results apropos of the charge densities and total charges that we deduced in the previous sections ensue here for $\bar{\rho}_c$.

We seek now for the distributional generalization of Eq.~(\ref{eqhydchargedTOV}). This can be easily done by following the same reasoning from the previous section, which finally leads us to
\begin{widetext}
\begin{equation}
\frac{d P}{dr} = \frac{Q(r)\bar{\rho}_c}{r^2} -{(\rho + P)}\alpha'_{Q} + \frac{2 {\cal P}_Q}{R}\delta(r-R)
+ \left\{[P(R)]^+_- + \frac{\sigma_Q}{2}[\alpha'_{Q+}(R) + \alpha'_{Q-}(R)] + \frac{\sigma_Q}{R} - \frac{[\alpha'_Q e^{-\beta_Q}]^+_-}{4\pi R} - \frac{\bar{\sigma}_{c}}{2R^2}\left[Q_+(R) + Q_-(R)\right] \right\}\delta(r-R)\label{hydrostaticeqQgr},
\end{equation}
\end{widetext}
where we are assuming surface quantities with the subindex ``$Q$'' are related to the charged versions of Eqs.~(\ref{enerdens}) and (\ref{surfdens}) [see also Eqs.~(\ref{ebetaq}) and (\ref{alphaprimeQ})]. It is easy to show that in the classical limit Eq.~(\ref{surfacetensionef}) naturally raises, implying that in such a limit the curly brackets in Eq.~(\ref{hydrostaticeqQgr}) is null.

We consider now the case where radial perturbations take place in our charged system. This case is more involved than the neutral case since the charged particles also feel an electric force. The equation describing the evolution of the displacements can be shown to be generalized to \citep[see][for the dynamics of the radial perturbations in a given phase]{2002PhRvD..65b4003A}
\begin{widetext}
\begin{eqnarray}
&&(\rho_0+P_0)e^{2(\beta_{Q0}-\alpha_{Q0})} \ddot{\tilde\xi}= -\frac{\partial P}{\partial r}- (\rho+P)\alpha'_{Q} + \frac{Q(r)\bar{\rho}_c}{r^2} + \frac{2{\cal P}_{Q}}{R} \delta(r-R)+ \bigg[ \frac{\sigma_{Q}}{R} - \frac{[\alpha'_{Q} e^{-\beta_{Q}}]^+_-}{4\pi R}  \nonumber \\ && + \frac{\sigma_{Q}}{2}\left\{e^{2(\beta_{Q0}^{+}-\alpha_{Q0}^{+})} \ddot{\tilde\xi}^{+} + e^{2(\beta_{Q0}^{-}-\alpha_{Q0}^{-})} \ddot{\tilde\xi}^{-} + \alpha'_{Q+} + \alpha'_{Q-}\right\}  -  \frac{\bar{\sigma}_{c}}{2R^2}\left[Q_+(R) + Q_-(R)\right] +  [P(R)]^+_- \bigg] \delta(r-R) \label{xiddotdistrQ}.
\end{eqnarray}
\end{widetext}
For the change in $Q(r)$, it can be shown \citep[see][]{1971PhRvD...4.2185B} that in the comoving frame there are no currents. This means that the Lagrangian displacements of $Q$ are null, $\Delta Q = 0$. Eq.~(\ref{xiddotdistrQ}) takes into account the values of the physical quantities in the presence of perturbations at $r$. In order to obtain the generalization of Eq.~(\ref{masterpertdistribsimpl}), we should evaluate Eq.~(\ref{xiddotdistrQ}) at $r+\tilde\xi$ and subtract it from Eq.~(\ref{hydrostaticeqQgr}). This is due to the definition of the Lagrangian displacement of a given physical quantity, intrinsically related to the notion of comoving observers with the fluid, who naturally could describe its thermodynamics.

In order to simplify Eq.~(\ref{xiddotdistrQ}), we have that in the generalized charged case \citep[see][for the
treatment in a phase of a charged system]{2002PhRvD..65b4003A}
\begin{widetext}
\begin{align}
\Delta \beta_Q &= -\alpha'_{Q0}\tilde\xi\label{dbetaQ},\\
\Delta\alpha'_Q &= 4\pi r e^{2\beta_{Q0}}\left[ \delta P + 2\delta \beta_Q \left(P_0-\frac{Q^2_0}{8\pi r^4} \right) \right] + \frac{e^{2\beta_{Q0}}}{r}\delta\beta_Q + \alpha'_{Q0}\tilde\xi + \frac{4\pi e^{2\beta_{Q0}}\bar{\rho}_c Q_0}{r}\tilde\xi  -\frac{2\pi\bar{\sigma}_c\tilde\xi}{R} \left(Q_0^+e^{2\beta_0^+} + Q_0^-e^{2\beta_0^-}\right) \delta(r-R) \label{dalphaprimeQ},\\
\Delta P &= -\gamma P_0 \left[ \frac{e^{-\beta_{Q0}}\left(r^2e^{\beta_{Q0}}\tilde\xi\right)'}{r^2} + \delta\beta_{Q} \right]\label{dpQ},\\
\Delta\rho &= -(\rho_0 + P_0) \left[\frac{1}{r^2}\frac{\partial}{\partial r}(r^2\tilde\xi) \right] -  \left[\frac{d P_0}{d r} - \frac{Q \bar{\rho}_c}{ r^2}\right]\tilde{\xi} + \bigg(\frac{\sigma_{Q0}}{2}\bigg\{\frac{\partial \tilde{\xi}^+}{\partial r} + \frac{\partial \tilde{\xi}^-}{\partial r} \bigg\} + \Delta\sigma_{Q} + \frac{2\sigma_{Q0}\tilde{\xi}}{R} + [P_0]^+_-\tilde{\xi} - \frac{\bar{\sigma}_{c}\tilde{\xi}}{2R^2}\left[Q_+ + Q_-\right]
\bigg)\delta(r-R)\label{drhoQ},\\
\gamma &\doteq \frac{\rho_0+ P_0}{P_0}\left(\frac{\partial P}{\partial \rho}\right)_{s={\rm const}}\label{gammaQ},\\
\Delta \sigma_{Q} &= -\left[\frac{2\sigma_{Q0}}{R} + [e^{\beta_{Q0}}P_0]^+_- + \frac{[\cosh\beta_{Q0}]^+_-}{4\pi R^2}\right] \tilde{\xi} \label{dsigmagrQ}.
\end{align}
\end{widetext}
The additional ``$0$'' subindex in a physical quantity means that its value at equilibrium was taken. We just stress that Eq.~(\ref{gammaQ}) is the general relativistic definition of the adiabatic index and assumes the existence of an equation of state
linking the pressure and the density of the system, $P=P(\rho)$. In this sense, it generalizes $\Gamma$ as defined by Eq.~(\ref{Gamma}).

By seeking for solutions $\tilde\xi = e^{i\omega t} \xi(r)$, one can see that Eq.~(\ref{xiddotdistrQ}) just gives meaningful boundary conditions when a frequency $\omega$ is the same for all the phases present in the system. We emphasize this is an universal property of the approach developed here, due to the surface degrees of freedom and the well-posedness of the problem of radial perturbations in stratified systems. The associated boundary condition rising from this analysis leads us to the conclusion that generically $\xi(r)$ is not differentiable at a surface of discontinuity, though continuous. It can be shown that the associated boundary condition to be taken into account here is functionally the same as Eq.~(\ref{deltatermgr}) [or Eq.~(\ref{deltatermgrsimp})], where now the metric and surface quantities should be related to the charged case.

\section{Conclusions}\label{sec:8}

In this article we have developed a formalism for assessing the stability of an stratified star against radial perturbations. We have derived the relevant equations defining this boundary-value problem for both neutral and charged stars and in Newtonian and Einstein's gravity. It makes use of the generalized theory of distributions, since we assumed that the surfaces of discontinuity are thin, that they host surface degrees of freedom, and that the phases separated by them do not mix. We showed that though the phases may be very different amongst themselves, when perturbations take place, they lead to the notion of a set of eigenfrequencies describing the whole system, instead of an independent set for each phase. Besides, our formalism gave us as a consequence the proper additional boundary conditions to be taken into account when working with stratified systems. Such boundary conditions encompass surface degrees of freedom in the surfaces of discontinuity and generically modify the set of eigenfrequencies with respect to their continuous counterpart. This should be a generic fingerprint of stratified systems with nontrivial surface degrees of freedom. Our analyses are relevant for the assessment of the stability of realistic star models, since they ensue the precise notion of boundary conditions. It was not our objective to systematically apply our formalism here, but simply to derive and expound it. It is clear that for precise and realistic numerical stability calculations, it would be ideal to have the microphysical knowledge of the properties of the interfacial surfaces. However, this a difficult problem which has been elusive even in the most advanced field of material science, where laboratory data of the material surface properties are accessible, but still with the lack of a complete physical theory for their explanation \citep{surfaceforces2011}. Thus, the measurement of the star's eigenmodes becomes of major relevance, since it could give information not only on the star's bulk structure but also on the possible existence of interior interfaces and their associate microphysical and electromagnetic phenomena.

The radial instabilities shown in our analyses should be interpreted analogously as for continuous stars since, even in the stratified case, a global set of eigenfrequencies raises. Thus, stratified stars would either implode or explode when they are radially unstable. Though we were concerned only with radial perturbations, it is of interest to investigate the additional oscillation modes owing to nonradial perturbations. The only point to be added, with respect to continuous stars, is the proper redefinition of the surfaces of discontinuity when such perturbations take place. This is clearly a richer scenario that insert additional degrees of freedom into the system, leading to the appearance of additional modes such as the gravitational `g-modes' \citep[see, e.g.,][]{1992ApJ...395..240R}. Such an analysis, however, is as a second step that goes beyond the goal of the present work, and that we are planning to investigate elsewhere.

\begin{acknowledgments}
We are grateful to Professor Thibault Damour for the discussions in the various occasions of the International Relativistic Astrophysics
(IRAP) PhD-Erasmus Mundus Joint Doctorate Schools held in Nice. We are likewise grateful to Professor Luis Herrera. J.P.P. acknowledges the support given by the Erasmus Mundus Joint Doctorate Program within the IRAP PhD, under the Grant Number 2011--1640 from EACEA of the European Commission.
J.A.R. acknowledges the support by the International Cooperation Program CAPES-ICRANet financed by CAPES -- Brazilian Federal Agency
for Support and Evaluation of Graduate Education within the Ministry of Education of Brazil.
\end{acknowledgments}


\end{document}